\documentclass[letterpaper,10pt]{article} 
\pdfoutput=1

\usepackage{opticameet3} 

\newcommand\authormark[1]{\textsuperscript{#1}}

\usepackage{amsmath,amssymb}
\usepackage[colorlinks=true,bookmarks=false,citecolor=blue,urlcolor=blue]{hyperref}
\usepackage{booktabs}
\usepackage{caption}
\usepackage{setspace}

\begin{document}

\title{Tensorized Optical Multimodal Fusion Network}


\author{Yequan Zhao\authormark{1}, Xian Xiao\authormark{2}, Geza Kurczveil\authormark{2}, Raymond G. Beausoleil\authormark{2}, and Zheng Zhang\authormark{1}}

\address{\authormark{1} Department of Electrical and Computer Engineering, University of California, Santa Barbara, CA 93106, USA\\
\authormark{2}Hewlett Packard Labs, Hewlett Packard Enterprise, 820 N. McCarthy Blvd., Milpitas, CA 95305, USA}
\email{yequan$\_$zhao@ucsb.edu, xian.xiao@hpe.com, zhengzhang@ece.ucsb.edu} 

\vspace{-10pt}
\begin{abstract}
We propose the first tensorized optical multimodal fusion network architecture with a self-attention mechanism and low-rank tensor fusion. Simulation results show 51.3$\times$ less hardware requirement and 3.7$\times$ 10$^{13}$ MAC/J energy efficiency.
\end{abstract}

\section{Introduction}
Modern machine learning (ML) applications often include heterogeneous data sources, e.g., visual, audio, and text data, in virtual reality (VR) interactions. Multimodal fusion networks offer a solution to robust performance under data uncertainty/attacks/corruptions by fully leveraging the complementary information from different modalities. However, modern edge devices' limited memory and power budget constrain their real-time performance. Optical ML accelerators, such as optical neural networks (ONNs), are competitive solutions for edge computing due to their wide bandwidth, high speed, and low energy consumption. Moreover, recent progress in tensorized optical neural networks (TONN) enables the implementation of a large-scale (e.g., 1024×1024) ONN with cascaded small-scale (e.g., 8×8) photonic tensor cores \cite{TONN}, which significantly improved the scalability of ONN.

In this paper, we propose a tensorized optical multimodal fusion network (TOMFN) to further explore TONN’s capability for processing large-scale and versatile ML data. By utilizing both tensor-train (TT) \cite{TTD} and CP decompositions \cite{CP}, the proposed compact and energy-efficient architecture can be implemented on integrated photonic platforms with small-scale photonic tensor cores as the building blocks. Moreover, compared to full-size counterparts, our proposed network maintains a compatible inference accuracy in multimodal sentiment analysis tasks while requiring 51.3× fewer hardware resources and excellent power reduction (i.e., 7.9nJ per inference).


\begin{figure}[!b]
\vspace{-10pt}
\centerline{\includegraphics[width=\textwidth]{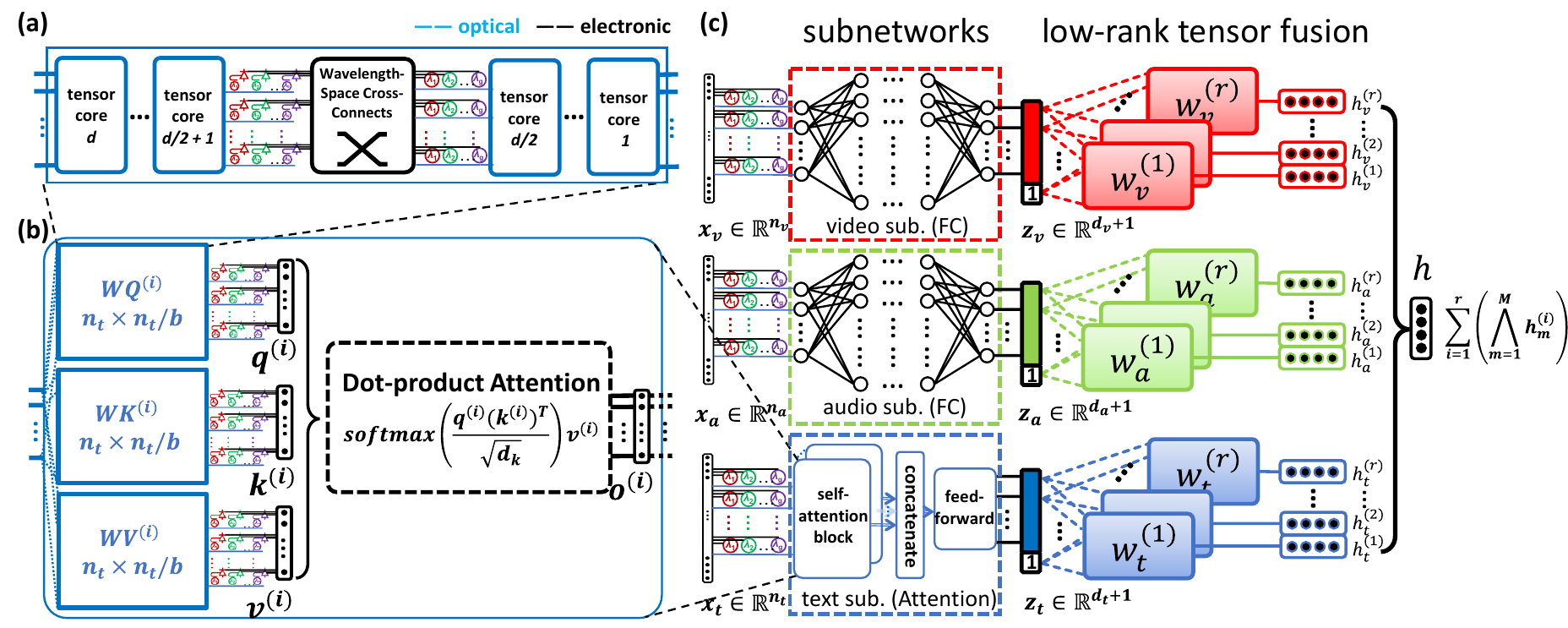}}
\captionsetup{font={footnotesize,stretch=1},justification=raggedright}
\caption{(a) Device implementation of a tensorized optical layer. 
(b) Detailed structure of a self-attention block. 
(c) Architecture of the tensorized optical multimodal fusion network (TOMFN).
}
\label{layer}
\end{figure}

\section{Principles and Architecture}
Fig.1 (c) depicts the conceptual architecture of the proposed TOMFN. It starts with three subnetworks that extract visual, audio, and text embeddings from corresponding input sources. The extracted vectors then go through a low-rank multimodal fusion network (LMF) \cite{LMF}, which maintains the expressiveness and effectiveness of tensor fusion network (TFN) \cite{TFN} while directly computing a rich and joint multimodal representation $h$ without explicitly creating the tensor. Lastly, the sentiment inference network consists of simply a softmax layer conditioned on $h$.

In particular, the processing of text embeddings is realized by the tensorized self-attention mechanism instead of the conventional LSTM layers. The benefit of self-attention is that it is a feed-forward structure without any recurrent layer, thus requiring no memory for intermediate states, and it has better parallelism. Fig. 1 (b) shows the detailed structure of the tensorized self-attention block. A single self-attention block computes one set of a dot-product attention score containing three optical tensorized layers that transform the input sequence $x_t$ into query vector $\mathbf{q}$, key vector $\mathbf{k}$, and value vector $\mathbf{v}$ in parallel. By concatenating multiple blocks, \textit{multi-head attention}, the model jointly attends to information from different representation subspaces at different positions \cite{Transformer}. The low-rank fusion network is also feed-forward. It comprises three modality-specific tensorized optical layers followed by element-wise product over the corresponding entries of each modality output and summation over fusion rank $r$. In practice, The TT and CP-format tensor ranks of all blocks could be learnt automatically in the training process~\cite{tt_training}. The element-wise product can be realized by two cascaded stages of balanced homodyne detection.

The photonic implementation of the large-scale weight matrices in fully-connected (FC) layers, fusion layers, and self-attention layers are all based on TONN architecture \cite{TONN}. TONN architecture takes advantage of the TT-decomposition algorithm and wavelength division multiplexing (WDM) technology, as shown in Fig. 1 (a). Such architecture can realize large-scale tensorized weight matrices by cascading small-radix photonic tensor cores implemented by Mach-Zehnder interferometer (MZI) meshes. This way, the number of MZIs and cascaded stages of MZIs are significantly reduced, leading to a significantly more compact footprint, smaller optical loss budget, and more manageable control complexity. More importantly, TONN architecture eliminates O/E/O conversions and intermediate memory, significantly improving energy efficiency and latency.

\section{Simulation Results and Conclusion}
The sentiment analysis result on the interactive emotional dyadic motion capture (IEMOCAP) dataset \cite{IEMOCAP} shows that our proposed TOMFN model has a competitive accuracy compared with the full-size LMF model. At the same time, our architecture exhibits 92.8× fewer model parameters, thus requiring 51.3× fewer photonic devices (i.e., MZIs). Note that our tensorized attention-based model outperforms its counterpart with a tensorized LSTM layer with the exact hardware requirements. In detail, we utilized three FC layers (80×32, 32×32, and 32×32) for the visual subnetwork, three FC layers (36×32, 32×32, and 32×32) for the audio subnetwork, and two blocks of self-attention layers (each contains three 300×150) and a feed-forward layer (300×64) for the text subnetwork. In total, thirty-eight 4×4, seventy-four 6×6, and sixteen 8×8 photonic tensor cores based on MZI meshes are needed for implementing all the weight matrices. Assuming the modulation speed is 10 Gb/s, the system power consumption is estimated at 79.87W, corresponding to 7.9 nJ per inference and 3.7×10$^{13}$ MAC/J.

\begin{table}[htbp]
  \centering
  \footnotesize
    \begin{tabular}{ccccccccc}
    \toprule
    \toprule
          & \multicolumn{3}{c}{Model size} & \multicolumn{4}{c}{IEMOCAP F1-result} & \multicolumn{1}{l}{Power estimation @10GHz} \\
\cmidrule{2-9}          & \# Param.$^*$ & \# MZI & \# stage & Happy & Sad   & Angry & Neutral & Efficiency \\
    \midrule
    TFN\cite{TFN} & 758176 & 607570 & 2841  & 83.6  & 82.8  & 84.2  & 65.4  & N/A \\
    LMF\cite{LMF} & 106912 & 86802 & 921   & \textbf{85.8} & \textbf{85.9} & \textbf{89} & \textbf{71.7} & N/A \\
    TOMFN(LSTM) & \textbf{844} & \textbf{1540} & 204   & 81.3  & 78.2  & 83.5  & 61.6  & 1.9$\times 10^{13}$ MAC/J \\
    TOMFN(Attention) & 1152  & 1691  & \textbf{166}   & \textbf{83.4} & \textbf{82.7} & \textbf{85.7} & \textbf{66.7} & \textbf{3.7$\mathbf{\times 10^{13}}$ MAC/J} \\
    \bottomrule
    \bottomrule
    \end{tabular}%
    \captionsetup{font={footnotesize,stretch=1},justification=raggedright}
    \caption{Test Accuracy and Power Consumption Analysis ($^*$ Only including the parameters of weight matrices)}
    \vspace{-10pt}
  \label{tab:addlabel}%
\end{table}%

In conclusion, given the success of attention-based architecture (e.g., Transformer\cite{Transformer}) in versatile applications, including natural language processing, object detection, and reinforcement learning, our proposed TOMFN lays the foundations for scalable, energy-efficient, and robust optical ML accelerators for heterougeneous data sources.



\end{document}